\documentclass[twocolumn,aps,english,prl,nofootinbib,preprintnumbers]{revtex4}
\usepackage{mathrsfs}
\usepackage{graphicx}
\usepackage{amsmath, amssymb}
\usepackage{babel}
\usepackage{color}
\usepackage{slashed}

\begin{document}
\preprint{ACFI-T18-12}
\preprint{MITP/18-070}

\title{Reduced hadronic uncertainty in the determination of $V_{ud}$}

\author{Chien-Yeah Seng$^{a}$}
\author{Mikhail Gorchtein$^{b}$}
\email{gorshtey@uni-mainz.de}
\author{Hiren H. Patel$^{c}$}
\author{Michael J. Ramsey-Musolf$^{c,d}$} 

\affiliation{$^{a}$INPAC, Shanghai Key Laboratory for Particle Physics and Cosmology, \\
	MOE Key Laboratory for Particle Physics, Astrophysics and Cosmology,  \\
	School of Physics and Astronomy, Shanghai Jiao-Tong University, Shanghai 200240, China}
\affiliation{$^{b}$Institut f\"ur Kernphysik, PRISMA Cluster of Excellence\\
	Johannes Gutenberg-Universit\"at, Mainz, Germany}
\affiliation{$^{c}$Amherst Center for Fundamental Interactions, Department of Physics, University of Massachusetts, Amherst, MA 01003}
\affiliation{$^{d}$Kellogg Radiation Laboratory, California Institute of Technology,
	Pasadena, CA 91125 USA}

\date{\today}

\begin{abstract}
We analyze the universal radiative correction $\Delta_R^V$ to neutron and superallowed nuclear $\beta$ decay by expressing the hadronic $\gamma W$-box contribution in terms of a dispersion relation, which we identify as an integral over the first Nachtmann moment of the $\gamma W$ interference structure function $F_3^{(0)}$.  By connecting the needed input to existing data on neutrino and antineutrino scattering, we obtain an updated value of $\Delta_R^V = 0.02467(22)$, wherein the hadronic uncertainty is reduced.  Assuming other Standard Model theoretical calculations and experimental measurements remain unchanged, we obtain an updated value of $|V_{ud}| = 0.97366(15)$, raising tension with the first row CKM unitarity constraint.  We comment on ways current and future experiments can provide input to our dispersive analysis.
\end{abstract}

\maketitle

The unitarity test of the Cabibbo-Kobayashi-Maskawa (CKM) matrix serves as one of the most important precision tests of the Standard Model.  In particular, tests of first-row CKM unitarity $|V_{ud}|^2+|V_{us}|^2+|V_{ub}|^2=1$ receive the most attention since these matrix elements are known with highest precision, all with comparable uncertainties.   The good agreement with unitarity  \cite{PDG2018} serves as a powerful tool to constrain New Physics scenarios.

Currently, the most precise determination of $|V_{ud}|$ comes from measurements of half-lives of superallowed $0^+\rightarrow0^+$ nuclear $\beta$ decays with a precision of $10^{-4}$ \cite{Hardy:2014qxa}.  At tree-level, these decays are mediated by the vector part of the weak charged current only, which is protected against renormalization by strong interactions due to conserved vector current (CVC), making the extraction of $|V_{ud}|$ relatively clean.  Beyond tree-level, however, electroweak radiative corrections (EWRC) involving the axial current are not protected, and lead to a hadronic uncertainty that dominates the error in the determination of $|V_{ud}|$. 

The master formula relating the CKM matrix element $|V_{ud}|$ to the superallowed nuclear $\beta$ decay half-life is \cite{Hardy:2014qxa}:
\begin{equation}
\left|V_{ud}\right|^2=\frac{2984.432(3)\,\text{s}}{\mathcal{F}t(1+\Delta_R^V)}\,,\label{eq:Vudsuper}
\end{equation}
where the nucleus-independent $\mathcal{F}t$-value is obtained from the experimentally measured $ft$-value by absorbing all nuclear-dependent corrections, and where $\Delta_R^V$ represents the nucleus-independent EWRC.  Currently, an average of the 14 best measured half-lives yields an extraordinarily precise value of $\mathcal{F}t=3072.27(72)$ s. A similar master formula exists for free neutron $\beta$ decay \cite{Czarnecki:2004cw} depending additionally on the axial-to-vector nucleon coupling ratio $\lambda=g_A/g_V$, and is free of nuclear-structure uncertainties.  But the much larger experimental errors in the measurement of its lifetime and the ratio $\lambda$ \cite{Czarnecki:2018okw} makes it less competitive in the extraction of $|V_{ud}|$.  Regardless, if first-row CKM unitarity is to be tested at a higher level of precision, improvement in the theoretical estimate of $\Delta_R^V$ by reducing hadronic uncertainties is essential.

The best determination of $\Delta_R^V = 0.02361(38)$ was obtained in 2006 by Marciano and Sirlin \cite{Marciano:2005ec} (in the following, we refer to their work as [MS]). They were able to reduce the hadronic uncertainty by a factor of 2 over their earlier calculation \cite{Marciano:1985pd} by using high order perturbative QCD corrections originally derived for the polarized Bjorken sum rule to precisely estimate the short distance contribution. At intermediate distances, an interpolating function motivated by vector meson dominance (VMD) was used to connect the long and short distances and was identified as the dominant source of theoretical uncertainty. This result leads to the current value of $|V_{ud}| = 0.97420(21)$ \cite{PDG2018}.

In this Letter, we introduce a new approach for evaluating $\Delta_R^V$ based on dispersion relations which relate it to directly measurable inclusive lepton-hadron and neutrino-hadron structure functions. 
Dispersion relations have proved crucial for evaluating the $\gamma Z$-box correction to parity violating electron-hadron interaction in atoms and in scattering processes \cite{Gorchtein:2008px,Sibirtsev:2010zg,Rislow:2010vi,Gorchtein:2011mz,Blunden:2011rd,Carlson:2012yi,Blunden:2012ty,Hall:2013hta,Rislow:2013vta,Hall:2013loa,Gorchtein:2015qha,Hall:2015loa,Gorchtein:2015naa}. It led to a significant shift in the 1-loop SM prediction for the hadronic weak charges, and ensured a correct extraction of the weak mixing angle at low energy \cite{Androic:2018kni}. 
Using existing data on neutrino and anti-neutrino scattering, we obtain a more precise value of the nucleus-independent EWRC,
\begin{equation}\label{eq:ourewrc}
\Delta_R^V = 0.02467(22)\,,
\end{equation} and therefore a new determination of $|V_{ud}|$,
\begin{equation}\label{eq:ourvud}
|V_{ud}| = 0.97366(15).
\end{equation}
We summarize in this Letter the essential features of our analysis that lead us to these values, and defer details to an upcoming longer paper \cite{Longpaper}.

\begin{figure}[t]
\includegraphics[width=0.7\columnwidth]{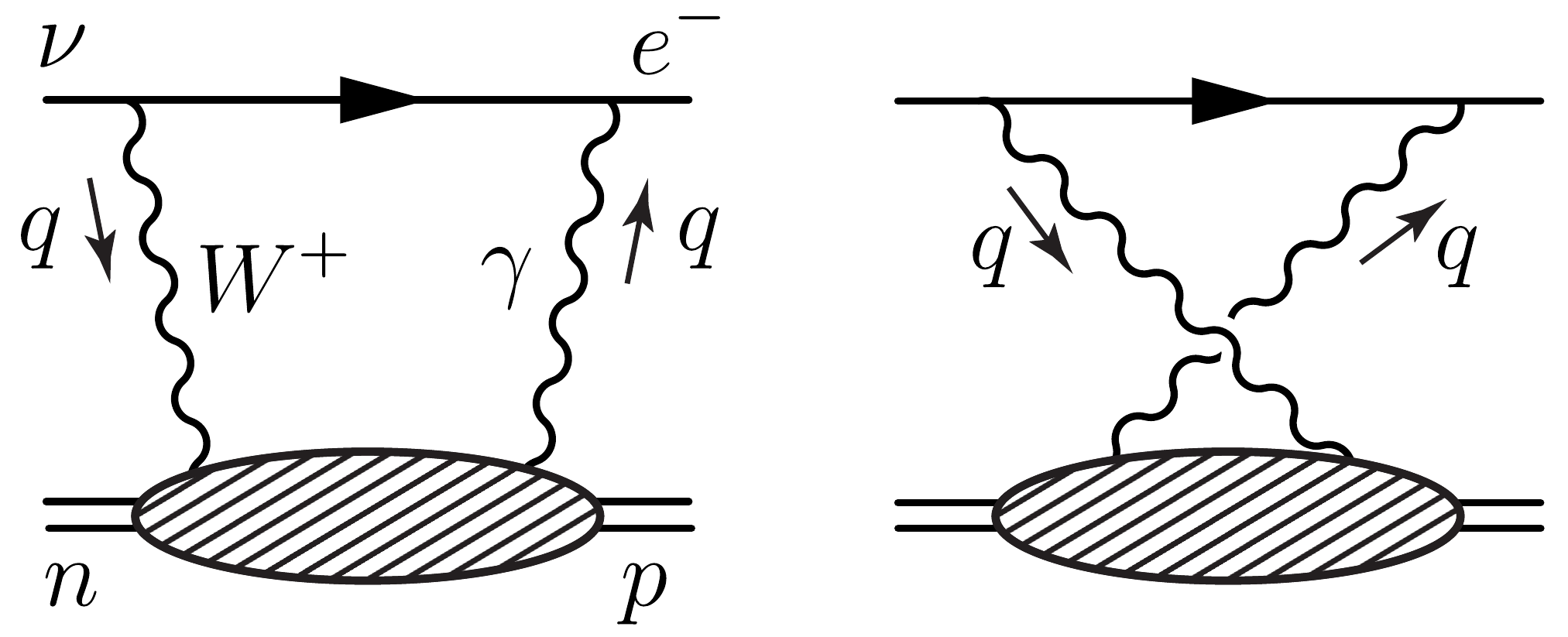}
\caption{Feynman diagrams corresponding to the amplitude in (\ref{eq:gaWloop}) which contribute at order $\mathcal{O}(\alpha/\pi)$ to neutron $\beta$ decay and are sensitive to the hadronic scale.}
\label{fig:feynmandiagrams}
\end{figure}

Among the various contributions at $\mathcal{O}(\alpha/\pi)$ to the neutron $\beta$ decay amplitude, Sirlin established \cite{Sirlin:1977sv} that the only one sensitive to the hadronic scale is the part in the $\gamma W$ box amplitude (Fig. \ref{fig:feynmandiagrams}),
\begin{multline}\label{eq:gaWloop}
\mathcal{M}_{VA}=2\sqrt{2}e^2G_FV_{ud}\int\frac{d^4q}{(2\pi)^4}\bigg[\\
\frac{\bar u_e(\mathbf{k})\gamma_\mu(\slashed{k}-\slashed{q}+m_e)\gamma_\nu P_L v_\nu(\mathbf{k})}{q^2[(k-q)^2-m_e^2]}\frac{M_W^2}{q^2-M_W^2}T^{\mu\nu}_{VA}\bigg],
\end{multline}
involving the nucleon matrix element of the product of the electromagnetic (EM) and the axial part of the weak charged current
\begin{equation}
T^{\mu\nu}_{VA} = \frac{1}{2}\int d^4x \, e^{iqx}\langle p(\mathbf{p})|T[J^\mu_\text{em}(x)J^{\nu}_{W,A}(0)]|n(\mathbf{p})\rangle\,.
\end{equation}
After inserting the nucleon matrix element parametrized in terms of the $P$-odd invariant function $T^{\mu\nu}_{VA} = \frac{i\epsilon^{\mu\nu\alpha\beta}p_\alpha q_\beta}{2 p\cdot q} T_3$ into the amplitude (\ref{eq:gaWloop}), the correction to the tree level amplitude is expressed as \cite{Sirlin:1977sv} 
\begin{multline}
\Box^{VA}_{\gamma W} = \frac{\alpha}{8\pi}\int_0^\infty dQ^2 \frac{M_W^2}{M_W^2+Q^2}\times\\
\int_{-i\sqrt{Q^2}}^{i\sqrt{Q^2}}\frac{d\nu}{\nu} \frac{4(Q^2+\nu^2)^{3/2}}{\pi M Q^4}T_3(\nu,Q^2)\label{eq:boxexplicit}
\end{multline}
where after Wick rotation the azimuthal angles of the loop momentum have been integrated over and the remaining integrals have been expressed in terms of $Q^2 = -q^2$ and $\nu=(p\cdot q)/M$.  With negligible error, we assume a common nucleon mass $M$ in the isospin symmetric limit and we work in the recoil-free approximation.  This contributes to the nucleus-independent EWRC as
\begin{equation}\label{eq:ewrc}
\Delta_R^V=2\Box_{\gamma W}^{VA}+\ldots\,,
\end{equation}
where the ellipses denote all other corrections insensitive to the hadronic scale.

Marciano and Sirlin estimate $\Box^{VA}_{\gamma W}$ by phenomenologically treating the $\nu$-integral $F_\text{M.S.}(Q^2) \equiv \int d\nu \ldots$ in the second line of (\ref{eq:boxexplicit}) as a function of $Q^2$, and parametrizing it piecewise over three domains: in the short distance domain $Q^2>(1.5\text{ GeV})^2$, the leading term in the OPE corrected by high order perturbative QCD is used; in the long distance domain $Q^2<(0.823\text{ GeV})^2$, the elastic nucleon with dipole form factors is used with a 10\% uncertainty; and at intermediate scales $(0.823\text{ GeV})^2<Q^2<(1.5\text{ GeV})^2$, an interpolating function inspired by VMD is used and is assigned a generous 100\% uncertainty.  Performing the integration over $Q^2$ in (\ref{eq:boxexplicit}) yields their value of $\Delta_R^V$ quoted above.

Our evaluation of $\Box^{VA}_{\gamma W}$ begins by first separating the invariant amplitude $T_3$ with respect to isosinglet and isotriplet components of the EM current $T_3=T_3^{(0)}+T_3^{(3)}$.  Crossing symmetry implies $T_3^{(0)}$ is odd under $\nu\rightarrow-\nu$ while $T_3^{(3)}$ is even.  Since the $\nu$ integration measure in (\ref{eq:boxexplicit}) is odd, only $T_3^{(0)}$ contributes to $\Box^{VA}_{\gamma W}$.  We then write a dispersion relation in $\nu$ for $T_3^{(0)}$, taking into account the physical sheet singularities.  Poles at $\nu_B=\pm Q^2/(2M)$ correspond to the elastic single-nucleon intermediate state, and branch points at $\nu_\pi=\pm(m_\pi^2+2Mm_\pi+Q^2)/(2M)$ correspond to single pion production thresholds.  We identify the discontinuity of $T_3^{(0)}$ across the cut as the $\gamma W$-interference structure function,  
$4\pi F^{(0)}_3(\nu,Q^2) = T_3^{(0)}(\nu+i\epsilon,Q^2)-T_3^{(0)}(\nu-i\epsilon,Q^2)$, so that the dispersion relation reads
\begin{equation}
T_3^{(0)}(\nu,Q^2)=-4i\nu\int_0^\infty d\nu'\frac{F_3^{(0)}(\nu',Q^2)}{\nu'^2-\nu^2}.\label{eq:DR}
\end{equation}
where $F_3^{(0)}$ contains both the elastic and inelastic contributions.  No subtraction constant appears since $T_3^{(0)}$ is an odd function of $\nu$.  Only $I=1/2$ intermediate states contribute because the EM current is isoscalar.  After inserting (\ref{eq:DR}) into (\ref{eq:boxexplicit}), performing the $\nu$-integral, and changing the integration variable $\nu' \rightarrow Q^2/(2Mx)$ we obtain
\begin{equation}
\Box^{VA}_{\gamma W}=\frac{3\alpha}{2\pi}\int_0^\infty \frac{dQ^2}{Q^2} \frac{M_W^2}{M_W^2+Q^2}M_3^{(0)}(1,Q^2),\label{eq:boxNachtmann}
\end{equation}
where $M_3^{(0)}(1,Q^2)$ is the first Nachtmann moment of the structure function $F_3^{(0)}$ \cite{Nachtmann:1973mr,Nachtmann:1974aj}
\begin{equation}
M_3^{(0)}(1,Q^2)=\frac{4}{3}\int_0^1 dx \frac{1+2r}{(1+r)^2}F_3^{(0)}(x,Q^2),
\label{eq:NachtmannDef}
\end{equation}
and $r=\sqrt{1+4M^2x^2/Q^2}$.  To estimate $\Box^{VA}_{\gamma W}$, we require the functional form of $F_3^{(0)}$ depending on $x$ and $Q^2$, or equivalently, $W^2=M^2+(1-x)Q^2/x$ and $Q^2$.

\begin{figure}[t]
\begin{center}
\includegraphics[width=0.9\columnwidth]{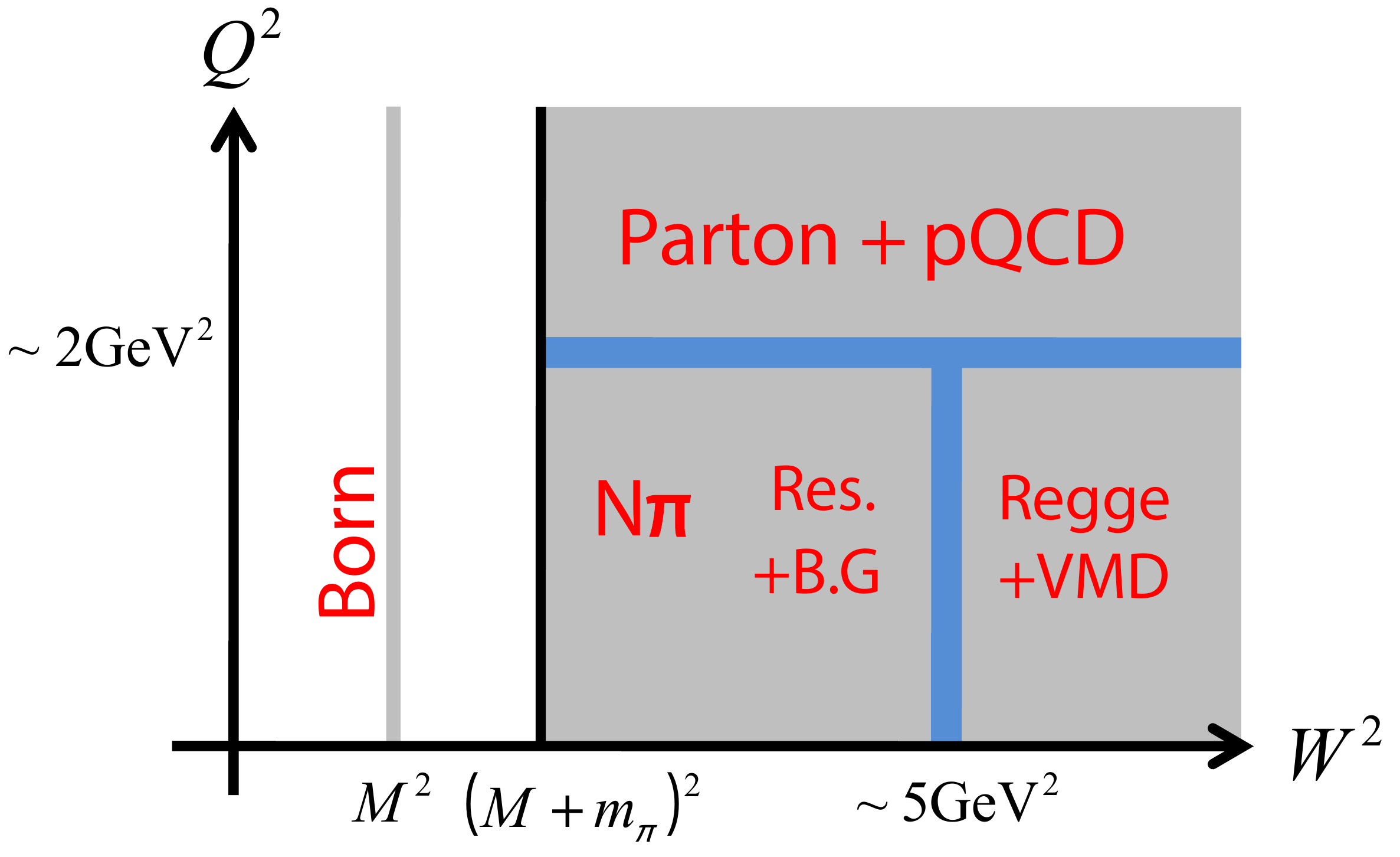}
\caption{Phase space of the structure functions $F_3^{(0)}$ and $F_3^{\nu p+\bar{\nu}p}$ in the $W^2$--$Q^2$ plane.}
\label{fig:W-Q2diag}
\end{center}
\end{figure}

We draw attention to the fact that (\ref{eq:boxNachtmann}) relates [MS]'s phenomenological function to the first Nachtmann moment
\begin{equation}\label{eq:msnachtmann}
F_\text{M.S.}(Q^2)=\frac{12}{Q^2}M_3^{(0)}(1,Q^2)\,,
\end{equation}
which will prove useful when comparing their results with ours.  Furthermore, since $F_3^{(0)}$ depends directly on on-shell intermediate hadronic states, it provides better handle on the physics that may enter at various scales.  Fig. \ref{fig:W-Q2diag} depicts the domain in the $W^2$--$Q^2$ plane over which $F_3^{(0)}$ has support: the single-nucleon elastic pole is at $W^2=M^2$, and the inelastic continuum covers the region above $W^2 > (M + m_\pi)^2$.

Our parameterization of $F_3^{(0)}$ is as follows:
\begin{equation}\label{eq:ourparam}
F_3^{(0)}\!\!=F_{\text{Born}}+
\begin{cases}
F_{\text{pQCD}}, & \!\! Q^2 \gtrsim 2\text{ GeV}^2 \\[2mm] 
\!F_{\pi N}\!+\!F_{\text{res}}\!+\!F_{\mathbb{R}}, & \!\!Q^2 \lesssim 2\text{ GeV}^2\,,
\end{cases}
\end{equation}
where each component is given by
\begin{gather}
\label{eq:bornpart}F_{\text{Born}}=-\frac{1}{4}(G_M^p+G_M^n)G_A\delta(1-x) \\
\label{eq:partonQCD}{\textstyle\int_0^1 dx\,F_{\text{pQCD}}} = \frac{1}{12}[1+\text{pQCD}] \\
\label{eq:threshold}F_{\pi N} = F_\text{$\chi$PT}\times(F_1^p+F_1^n)\frac{|G_A|}{g_A} \\
F_\text{res} = \text{negligible} \\
\label{eq:regge}F_{\mathbb{R}} = C_{\gamma W}f_\text{th}\frac{m_\omega^2}{m_\omega^2+Q^2}\frac{m_{a_1}^2}{m_{a_1}^2+Q^2}\left(\frac{\nu}{\nu_0}\right)^{\alpha_0^\rho}\,,
\end{gather}
and supplies the dominant contribution to $F_3^{(0)}$ in various regions indicated in Fig.~\ref{fig:W-Q2diag} which we describe next.

We obtain the elastic Born contribution at $W^2 = M^2$ in (\ref{eq:bornpart}) by using the updated values of the magnetic Sachs form factor $G_M$ and the axial form factor $G_A$ for the nucleon \cite{Ye:2017gyb,Bhattacharya:2011ah}.  Above threshold, $W^2 \geq (M+m_\pi)^2$, we consider the dominant physics operating in various of domains in the $Q^2$--$W^2$ plane separately. At large $Q^2 \gtrsim 2\text{ GeV}^2$, the Nachtmann moment $M_3^{(0)}$ reduces to the Mellin moment and is fixed by the sum rule corrected by pQCD in Eq. (\ref{eq:partonQCD}) by analogy with that of the polarized Bjorken sum rule [MS].  At small $Q^2 \lesssim 2\text{ GeV}^2$, we estimate the contribution (\ref{eq:threshold}) near the inelastic threshold by computing the single pion production contribution  $F_{\chi\text{PT}}$ in Chiral Perturbation Theory ($\chi$PT) at leading order.  To improve the behavior of $F_{\chi\text{PT}}$ at larger $Q^2$, we replace the point-like nucleon vertices with measured Dirac and axial nucleon form factors, $F_1$ and $G_A$.  At higher $W^2$, we investigated the impact of several low-lying $I=1/2$ resonances based on a few models \cite{Lalakulich:2006sw,Drechsel:2007if,Tiator:2008kd}, and found their contributions to $\Box_{\gamma W}^{VA}$ to be negligible.

\begin{figure}[t]
\begin{center}
\includegraphics[width=0.8\columnwidth]{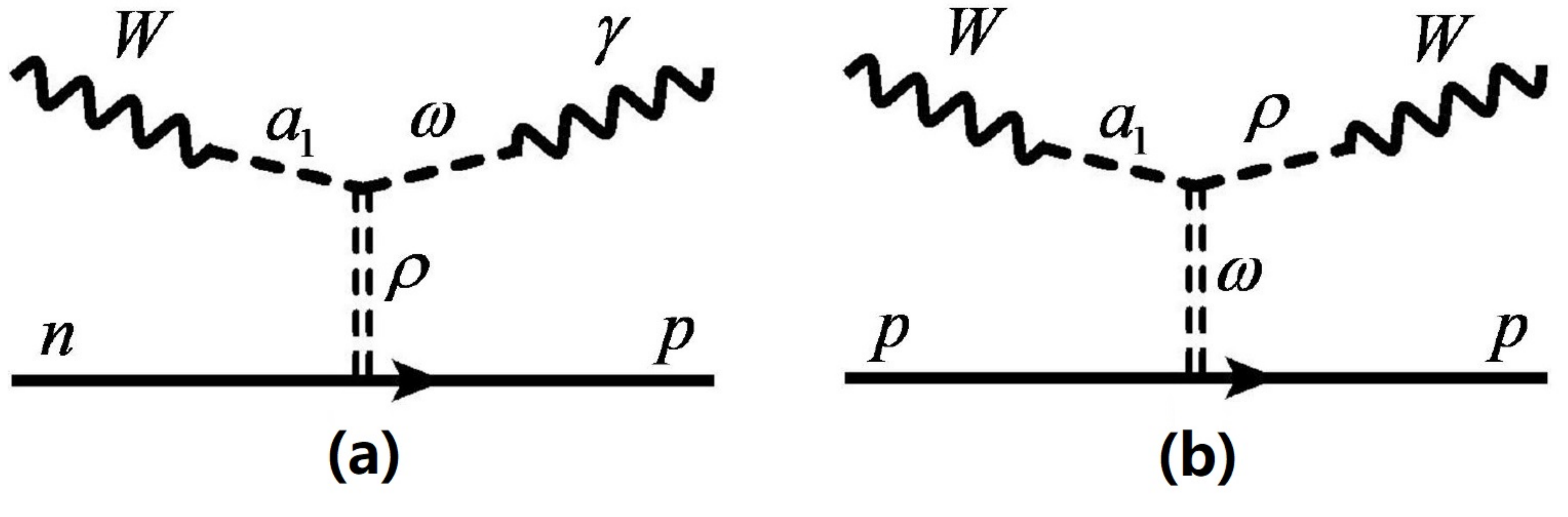}
\caption{Regge exchange model (a) for $F_3^{(0)}$ and (b) for $F_3^{\nu p+\bar{\nu}p}$ using vector meson dominance.}\label{fig:Regge}
\end{center}
\end{figure}

Finally, at large $W^2$, we use the form in Eq. (\ref{eq:regge}) inspired by Regge phenomenology together with VMD \cite{Piketty:1970sq} as illustrated in Fig.~\ref{fig:Regge}a.  In this picture, the Regge behavior $(\nu/\nu_0)^{\alpha_0^\rho}$ arises from the exchange of the $\rho$ trajectory with intercept $\alpha_0^\rho = 0.477$ \cite{Kashevarov:2017vyl}, and is coupled to the external currents via $a_1$ and $\omega$ mesons encoded by the VMD factors $m^2_V/(m_V^2 + Q^2)$.  We include a threshold function $f_\text{th}=\Theta(W^2-W_\text{th}^2)\left(1-\exp[(W_\text{th}^2-W^2)/\Lambda_\text{th}^2]\right)$ which smoothly vanishes at the two-pion threshold point $W_\text{th}^2=(M+2m_\pi)^2$ to model the smooth background in the resonance region \cite{Gorchtein:2011mz}.  We choose equal values for the Regge and threshold scales of $\nu_0 = \Lambda_\text{th} = 1\text{ GeV}$, to ensure that Regge behavior sets in around $W^2\sim(2.5\text{ GeV})^2$.  The function $C_{\gamma W} (Q^2)$ accounts for residual $Q^2$-dependence beyond that of the VMD, which we infer from experimental data as explained below.

\begin{figure}[h]
\begin{center}
\includegraphics[width=0.9\columnwidth]{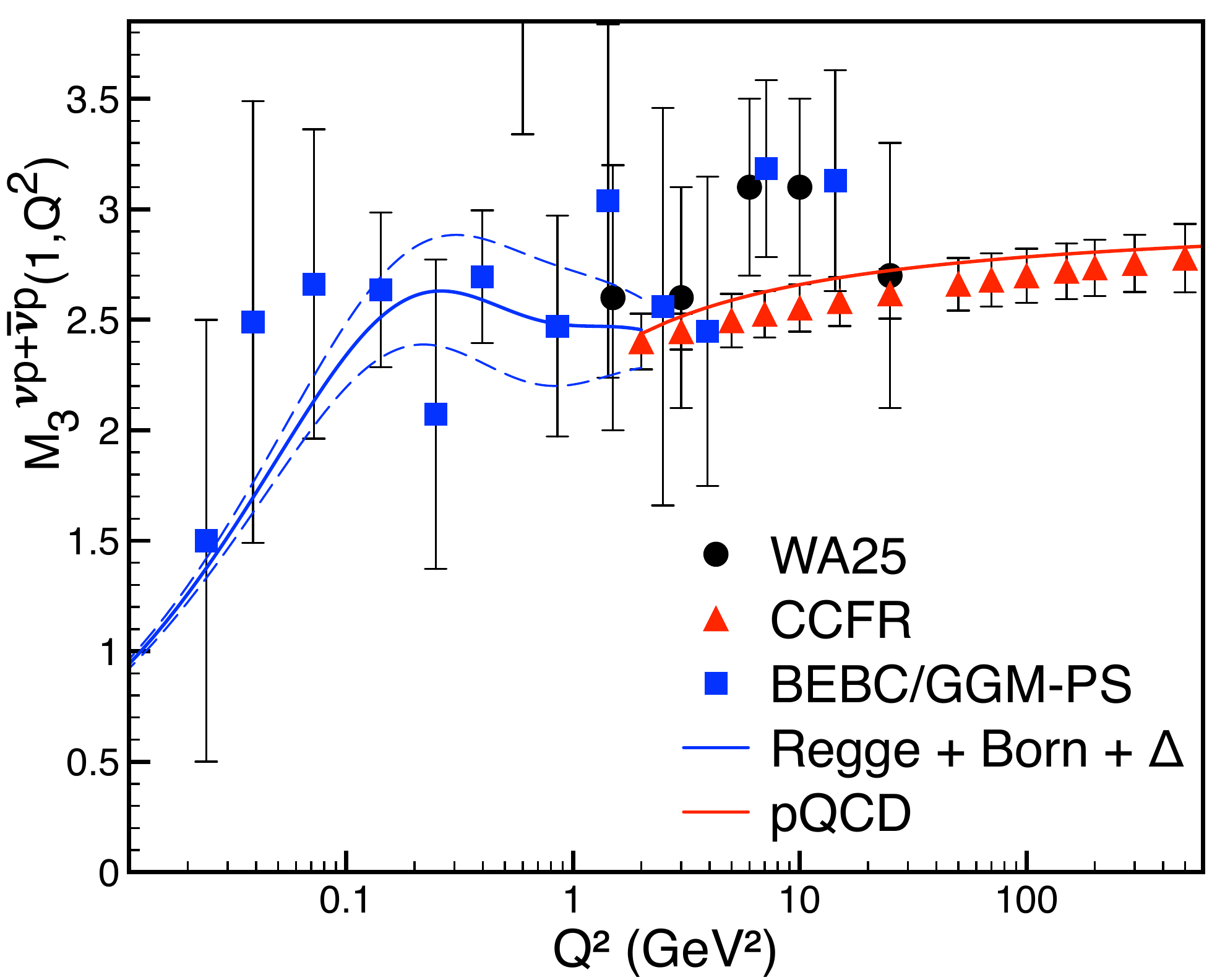}
\caption{World data of the first Nachtmann moment $M_{3}^{\nu p+\bar{\nu}p}(1,Q^2)$.  The red curve is the pQCD-corrected GLS sum rule above $Q^2\approx2\text{ GeV}^2$, and the blue curve is the result of the fit for $A_{WW}$ and $B_{WW}$ in (\ref{eq:linear}).}
\label{fig:GLSSJune29mod}
\end{center}
\end{figure}

Since the isospin structure of $F_3^{(0)}$ is \mbox{$(I=0)\times(I=1)$}, it is not directly accessible experimentally.  However, information about the $P$-odd structure function with a different isospin structure \mbox{$(I=1)\times(I=1)$} is available from $\nu$- and $\bar\nu$-scattering.  In particular, data exists on the first Nachtmann moment $M_3^{\nu p+\bar\nu p}$ for the combination $F_{3}^{\nu p+\bar{\nu}p}=(F_3^{W^-}+F_3^{W^+})/2$ derived from the difference of $\nu p$ and $\bar \nu p$ differential cross sections.  The data by CCFR \cite{Kataev:1994ty,Kim:1998kia}, BEBC/Gargamelle \cite{Bolognese:1982zd} and WA25 \cite{Allasia:1985hw} cover a wide region of $Q^2$ from 0.15 to 600 GeV$^2$ (see Fig. \ref{fig:GLSSJune29mod}).  Although the precision below $Q^2\approx1.4\text{ GeV}^2$ is less satisfactory, we are able to use it to collect information about the form of the analogous Regge coefficient function $C_{WW}(Q^2)$ for this structure function, and thereby infer the form of the required $C_{\gamma W}(Q^2)$ as follows. 

We parametrize the structure function $F_{3}^{\nu p+\bar{\nu}p}$ in precisely the same way as in (\ref{eq:ourparam}) for $F_3^{(0)}$, and establish $F_\text{Born}^{\nu p+\bar{\nu}p}$, $F_\text{pQCD}^{\nu p+\bar{\nu}p}$, $F_{\pi N}^{\nu p+\bar{\nu}p}$, $F_\text{res}^{\nu p+\bar{\nu}p}$ and $F_\mathbb{R}^{\nu p+\bar{\nu}p}$ along similar lines.  In this case, $\int_0^1 dx\, F_\text{pQCD}^{\nu p+\bar{\nu}p}$ satisfies the Gross-Llewellyn-Smith sum rule \cite{Gross:1969jf} corrected by pQCD \cite{Larin:1991tj}, while at low $Q^2$, the $\Delta$-resonance and the Born contribution saturate the Nachtmann moment \cite{Bolognese:1982zd}.  At large $W^2$, the $\omega$ trajectory controls the leading behavior, and couples to the external currents by the $a_1$ and $\rho$ mesons (see Fig. \ref{fig:Regge}b), leading to
\begin{equation}
F_{\mathbb{R}}^{\nu p+\bar{\nu}p}=C_{WW}f_\text{th}\frac{m_\rho^2}{m_\rho^2+Q^2}\frac{m_{a_1}^2}{m_{a_1}^2+Q^2}\left(\frac{\nu}{\nu_0}\right)^{\alpha_0^\omega}\,.
\end{equation}
We then fit the unknown function $C_{WW}(Q^2)$ to the data for $M_3^{\nu p+\bar{\nu}p}(1,Q^2)$ in the range $Q^2\leq 2$ GeV$^2$. Due to the quality of the data, we choose the simple linear form
\begin{equation}\label{eq:linear}
C_{WW}(Q^2)=A_{WW}(1+B_{WW}Q^2)
\end{equation}
and obtain $A_{WW}=5.2\pm 1.5$, $B_{WW}=1.08^{+0.48}_{-0.28}\text{ GeV}^{-2}$.  The result of the fit is shown by the blue curve in Fig. \ref{fig:GLSSJune29mod}.  The solid curve corresponds to the central value of the fit, and the dotted curve indicates the maximum variation in $M_3^{\nu p+\bar{\nu}p}$ allowed by the errors in the fit. We do not fit the three data points below $Q^2=0.1$ GeV$^2$ where Born and resonance contributions dominate the GLS sum rule: rather, we use the resonance parameters obtained in \cite{Lalakulich:2006sw} from a fit to modern neutrino data.

Finally, to obtain $C_{\gamma W}(Q^2)$, we require the ratio of Nachtmann moments $M_{3,\mathbb{R}}^{(0)}(1,Q^2)/M_{3,\mathbb{R}}^{\nu p+\bar{\nu}p}(1,Q^2)$ to agree with the value predicted by VMD at $Q^2=0$, and the QCD-corrected parton model at $Q^2=2 \text{ GeV}^2$. Since the $\rho$ and $\omega$ Regge trajectories are nearly degenerate \cite{Kashevarov:2017vyl}, the two conditions predict the same ratio \cite{Longpaper}
\begin{equation}
\frac{M_{3,\mathbb{R}}^{(0)}(1,0)}{M_{3,\mathbb{R}}^{\nu p+\bar{\nu}p}(1,0)}\approx\frac{M_{3,\mathbb{R}}^{(0)}(1,2\text{ GeV}^2)}{M_{3,\mathbb{R}}^{\nu p+\bar{\nu}p}(1,2\text{ GeV}^2)}\approx\frac{1}{36}.\label{eq:matching}
\end{equation}
For the linear parametrization in Eq.~(\ref{eq:linear}), this implies 
\begin{equation}\label{eq:reggecoeff}
C_{\gamma W}(Q^2)=\frac{1}{36}C_{WW}(Q^2)\,,
\end{equation}
providing us with the final piece of $F_\mathbb{R}$ in (\ref{eq:regge}).

Upon inserting our parameterization (\ref{eq:ourparam}) for the structure function $F_3^{(0)}$ into (\ref{eq:boxNachtmann},~\ref{eq:NachtmannDef}) and performing the integrations, we obtain the following contributions to $\Box_{\gamma W}^{VA}$ in units of $10^{-3}$: $2.17(0)$ from parton+pQCD, $1.06(6)$ from Born and $0.56(8)$ from Regge+resonance+$\pi N$, the digit in parentheses indicating the uncertainty. Combining them with the remaining known contributions [MS] gives our new values, 
$\Delta_R^V=0.02467(22)$ and $|V_{ud}|=0.97366(15)$. Our reevaluation of $\Delta_R^V$ represents a reduction in theoretical uncertainty over the previous [MS] result by nearly a factor of 2.  However, it also leads to a substantial upward shift in the central value of $\Delta_R^V$ and a corresponding downward shift of $|V_{ud}|$ by nearly three times their quoted error, now raising tension with the first-row CKM unitarity constraint: $|V_{ud}|^2+|V_{us}|^2+|V_{ub}|^2=0.9983(4)$.

\begin{figure}[h]
\begin{center}
\includegraphics[width=0.9\columnwidth]{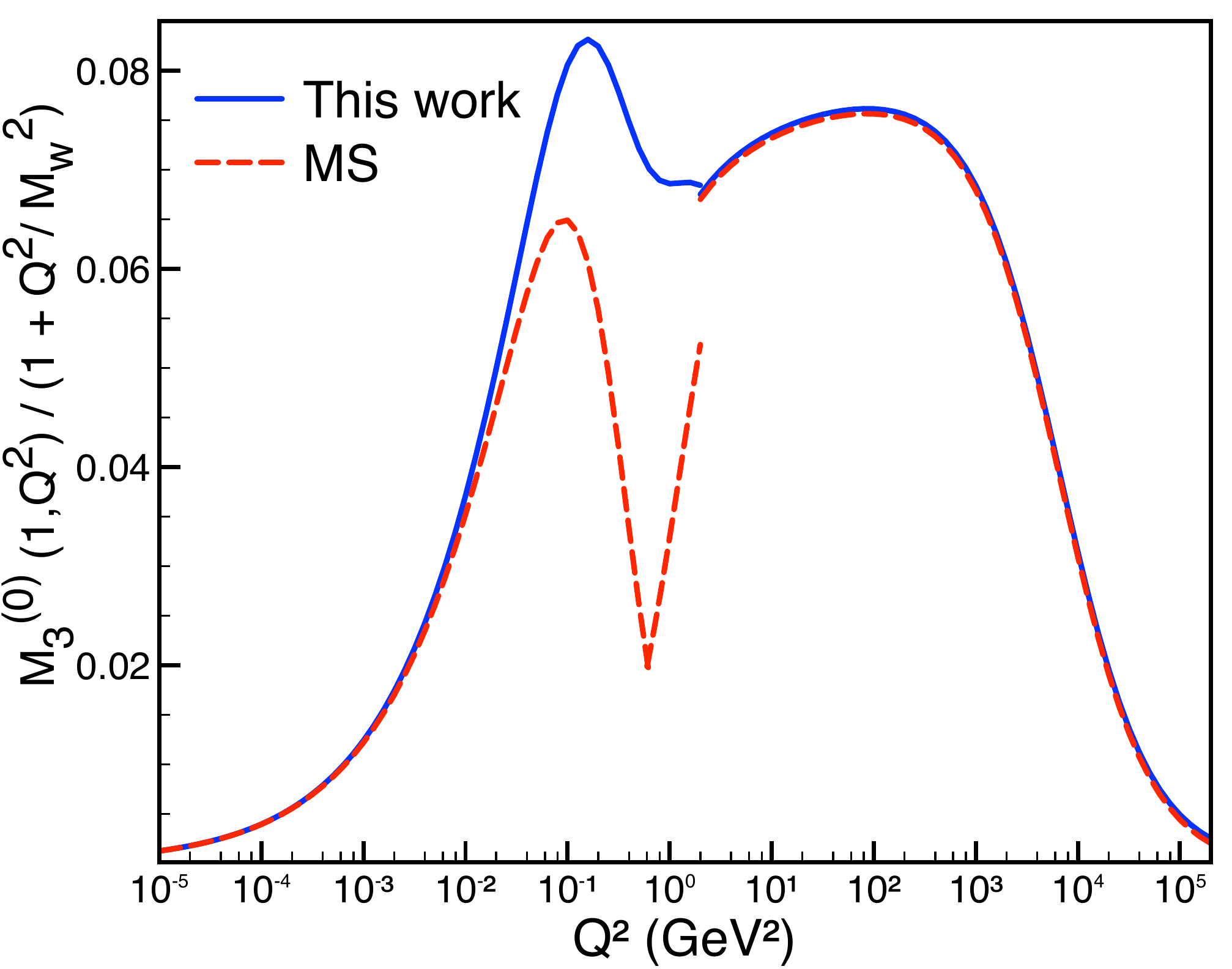}
\caption{Log-linear plot of $\frac{M_W^2}{M_W^2+Q^2} M_3^{(0)}(1,Q^2)$ as a function of $Q^2$.  The blue curve is the result of our parameterization in (\ref{eq:ourparam}), and the red curve is the piecewise parametrization used by [MS].  For a given parametrization, the contribution to $\Box^{VA}_{\gamma W}$ is proportional to the area under the curve, see (\ref{eq:boxNachtmann}). 
}
\label{fig:M3plot}
\end{center}
\end{figure}

We pause to comment on the origin of the large shift in the central value for $\Delta_R^V$ with respect to [MS].  In Fig. \ref{fig:M3plot} we plot the integrand $\frac{M_W^2}{M_W^2 + Q^2} M_3^{(0)}(1,Q^2)$ of Eq. (\ref{eq:boxNachtmann}) as a function of $Q^2$.  In solid blue, we show the result of our parametrization (\ref{eq:ourparam}) after integrating over $x$.  In dashed red, we show the piecewise parametrization by [MS] obtained with the help of (\ref{eq:msnachtmann}). 
The discontinuity in their parametrization at $Q^2 = (1.5\text{ GeV})^2$ arises from their choice of matching the $Q^2$ integrals of pQCD and the interpolating function over the short distance domain, rather than matching the functions themselves.
The log-linear scale conveniently accounts for the integration measure $dQ^2/Q^2$ in (\ref{eq:boxNachtmann}) so that the correction $\Box^{VA}_{\gamma W}$ is directly proportional to the area under the curve.  Although the line shapes are in agreement above $Q^2\gtrsim2\text{ GeV}^2$ and below $Q^2\lesssim0.001\text{ GeV}^2$, ours lies significantly above that of [MS] for intermediate $Q^2$. This difference is the origin of the discrepancy between our central values for $\Delta_R^V$.  By working with the two-variable structure function $F_3^{(0)}(x, Q^2)$, we were able to capture a broad variety of physics (Born, $N\pi$, Regge) operating at intermediate $Q^2$ in contrast with the one-variable analysis of $F_\text{M.S.}(Q^2)$ by [MS]. 
We therefore believe our updated result provides a more realistic assessment of $\Delta_R^V$, even though the difference with them is larger than their quoted theoretical uncertainty.

We conclude by discussing how new measurements could provide tests of our parameterization of $F_3^{(0)}$ and further reduce the uncertainty in $\Delta_R^V$. In view of the upcoming high-intensity neutrino beam program at Fermilab, we wish to point out the potential impact which new, more precise measurements of $M_3^{\nu p+\bar{\nu}p}(1,Q^2)$ at low $Q^2$ can have on our fit, as evidenced by Fig. \ref{fig:GLSSJune29mod}. That said, we have related $F_3^{(0)}$ and $F_{3}^{\nu p+\bar{\nu}p}$ within a model. However, by making use of isospin symmetry, we can establish a more robust relationship between $F_3^{(0)}$ and the $P$-odd structure function $F_{3,\gamma Z}^{N}$. The latter is accessible with parity-violating deep inelastic (inclusive) electron scattering. Since the axial component of the weak neutral current is predominantly isovector, we obtain
\begin{equation}
4F_3^{(0)}\approx F_{3,\gamma Z}^p-F_{3,\gamma Z}^n\approx 2F_{3,\gamma Z}^p-F_{3,\gamma Z}^d.\label{eq:gamamZisospin}
\end{equation}
\indent
Thus, fixed target measurements using hydrogen and deuterium can in principle provide a more direct way to determine $\Box_{\gamma W}^{VA}$ from data.  High quality data in the range $0.1\text{ GeV}^2 \lesssim Q^2 \lesssim 1\text{ GeV}^2$ and $W^2 \gtrsim 5\text{ GeV}^2$ would be particularly advantageous, as our parametrization of $F_3^{(0)}$ admits the greatest model-dependence and exhibits the largest difference from that of [MS] in this domain.  Such an experimental program will however require a dedicated feasibility study, as the contribution of $F_{3,\gamma Z}$ to the parity-violating asymmetry with a polarized electron beam is suppressed by the small weak charge of the electron.  Finally, with the  reduction in the uncertainty of $|V_{ud}|$, the error in the first-row CKM unitarity constraint is dominated by the uncertainty in $|V_{us}| = 0.2243(5)$. Combined with our results presented here, a commensurate reduction in the latter uncertainty would enhance the impact of first row CKM unitarity tests.

We acknowledge helpful discussions with Bill Marciano, John Hardy, Vincenzo Cirigliano, Peter Blunden, Emmanuel Paschos, Cheng-Pang Liu, Chung-Wen Kao, Hubert Spiesberger and Jens Erler.  Significant progress was made during the scientific program ``Bridging the Standard Model to New Physics with the Parity Violation Program at MESA" hosted by MITP Mainz. 
M.G.'s work was supported by the Deutsche Forschungsgemeinschaft under the personal grant GO 2604/2-1. CYS's work is supported in part by the National Natural Science Foundation of China (NSFC) under Grant Nos.11575110, 11655002, 11735010, Natural Science Foundation of Shanghai under Grant No.~15DZ2272100 and No.~15ZR1423100, by Shanghai Key Laboratory for Particle Physics and Cosmology, and by Key Laboratory for Particle Physics, Astrophysics and Cosmology, Ministry of Education, and also appreciates the support through the Recruitment Program of Foreign Young Talents from the State Administration of Foreign Expert Affairs, China.  HHP and MJRM were supported in part by US Department of Energy Contract DE-SC0011095.

%\bibliography{gaW_ref_letter}

\end{document}